\documentclass[twocolumn,english,APS, prl]{revtex4-1}
\usepackage[T1]{fontenc}
\usepackage[latin9]{inputenc}
\usepackage{color}
\usepackage{amsmath}
\usepackage{amssymb}
\usepackage{graphicx}
\usepackage{setspace}

\makeatletter

\newcommand*\LyXThinSpace{\,\hspace{0pt}}

\DeclareMathOperator{\sech}{sech}

\makeatother

\usepackage{babel}
\begin{document}

\title{Pulsed entanglement of two optomechanical oscillators}

\author{S. Kiesewetter, R. Y. Teh, M. D. Reid and P. D. Drummond}

\affiliation{Center for Quantum and Optical Science, Swinburne University of Technology,
Melbourne, Australia}
\begin{abstract}
A strategy for generating entanglement in two separated optomechanical
oscillators is analysed, using entangled radiation produced from downconversion
and stored in an initiating cavity. We show that the use of pulsed
entanglement with optimally shaped temporal modes can efficiently
transfer quantum entanglement into a mechanical mode, then remove
it after a fixed waiting time for measurement. This protocol could
provide new avenues to test for bounds on decoherence in massive systems
that are spatially separated, as originally suggested by Furry \cite{furry}
not long after the discussion by Einstein-Podolsky-Rosen (EPR) and
Schr\"odinger of entanglement.
\end{abstract}
\maketitle
Macroscopic mechanical oscillators have now been cooled to their quantum
ground state \cite{cooling_Connell_2010_Nature,cooling_Chan_2011_Nature,cooling_Gro_2009_NP,cooling_Teufel},
followed by the observations of macroscopic quantum effects \cite{quantum_macro,quantum_macro2,quantum_macro3,quantum_macro4}
including quantum entanglement \cite{Palomaki} between a mechanical
oscillator and a radiation field. Even more spectacular demonstrations
of macroscopic quantum properties will soon become achievable. An
important goal is to demonstrate long-lived entanglement between two
separated mechanical systems. This would enable new tests of quantum
mechanics, including the possible decoherence of quantum correlations
due to space-like separation. 

The intriguing idea of spatially dependent decoherence \cite{furry}
was first advanced by Furry just after the publication of the original
EPR paradox \cite{EPR1935} and entanglement papers \cite{Schrodinger}.
Furry's hypothesis is not predicted by conventional reservoir theory.
It could occur in a modified quantum mechanics, including quantum
gravity or other type of intrinsic decoherence \cite{Pearle,GRW,Penrose,Diosi}.
It is clear from experiments in quantum optics that spatially dependent
decoherence does not occur for massless photons \cite{Delft_2015}.
However, there are no measurements yet on entanglement decay when
massive, separated objects have an entangled center-of-mass motion.
The success of quantum optomechanical entanglement and gravity-wave
detectors \cite{LIGO_2013} demonstrates that such experiments are
ideal for investigating previously accessible questions like this.

\begin{figure}
\centering{}\includegraphics[width=0.8\columnwidth]{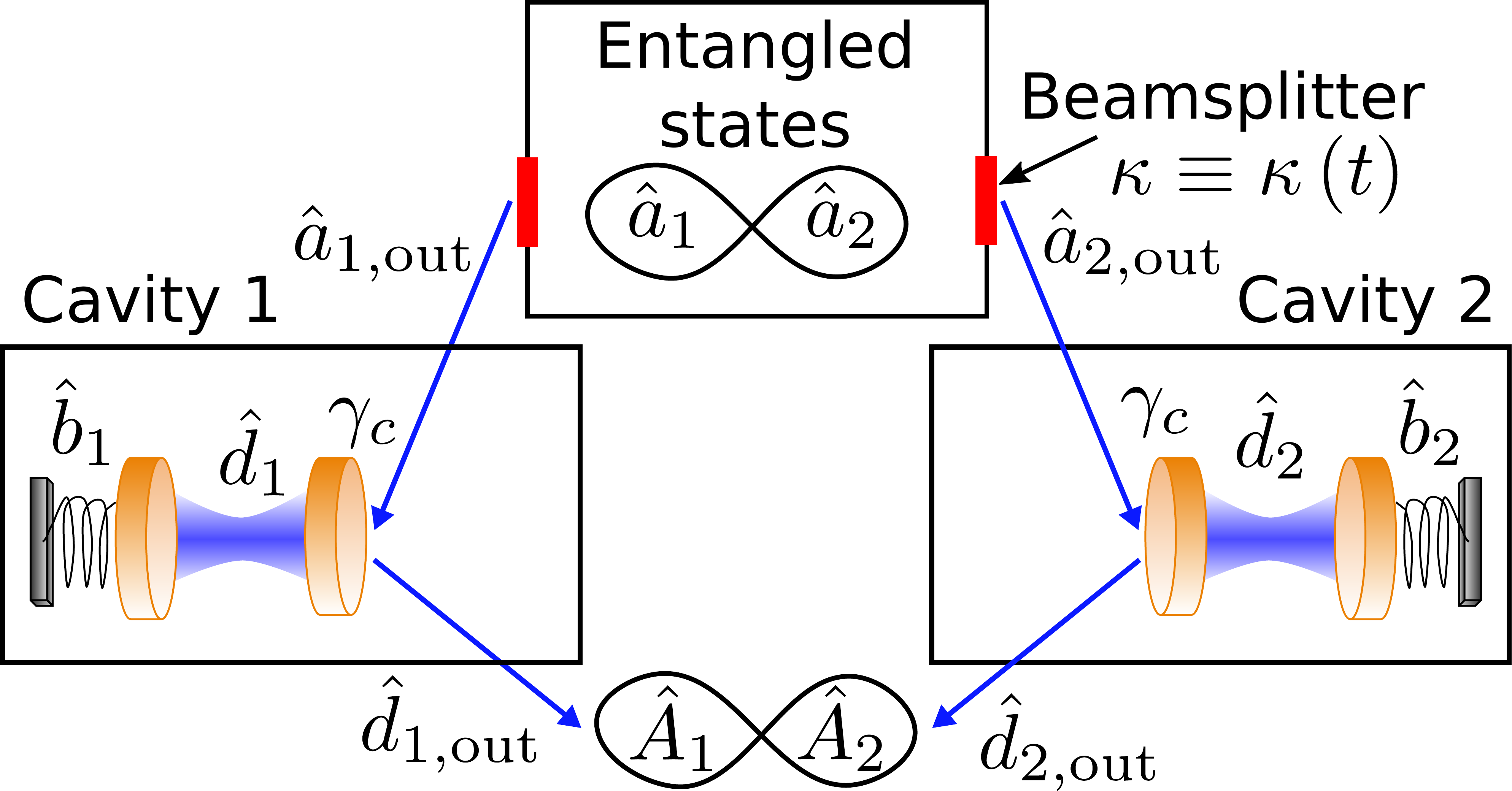}\caption{Schematic diagram of entanglement protocol.\label{fig:Schematic-diagram-of}}
\end{figure}

In this Letter, we propose and analyse a simple pulsed protocol for
creating and measuring such macroscopic entanglement. The basic experimental
setup involves an entangled source and spatially separated quantum
optomechanical systems. An optical parametric amplifier creates two
entangled modes \cite{Reid1989,Reid-Drumm,EPR-exp,rmp}, ideally with
the same frequency and different polarizations. This entanglement
is transferred, on demand, to the separated quantum optomechanical
systems \textendash{} thus destroying the initial entanglement in
optical modes. The entangled mechanical modes are stored, subsequently
coupled out and measured optically, as indicated schematically in
Fig (\ref{fig:Schematic-diagram-of}). There are other proposals for
entangling quantum optomechanical systems \cite{ent-opto-Giovannetti,ent-opto-Mancini,ent-opto-Hofer,ent_proposal-eanglement-criteria-He2013},
but the lifetime of quantum entanglement and spatial separation are
not controllable with these proposals. These requirements appear essential
to a test of Furry's hypothesis.

The entangled source cavity modes $a_{1}$ and $a_{2}$ are assumed
to be initially in a two mode squeezed state, prepared using the standard
technique of nondegenerate parametric down-conversion. To give a definite
model, this entangled state is initially prepared in a source cavity
whose entanglement is characterized by a squeezing parameter $r$.
The source cavity has tunable decay rates $\kappa\left(t\right)$,
generating shaped, entangled outputs \cite{sech_pulse_He2009}; other
methods for state preparation are also possible. 

The outputs are fed \cite{cascade_kolobov_1987,cascade_carmichael_1993,cascade_gardiner_1993}
into the quantum optomechanical systems labelled Cavity $1$ and Cavity
$2$, respectively, assuming identical optomechanical parameters.
We linearize the equations of motion for an adiabatic \cite{ent-opto-Giovannetti,ent-opto-Mancini}
pulsed optomechanical Hamiltonian describing these devices \cite{fast-coupling-Vanner2011,ent-opto-Hofer,fast-coupling-Pikovski-np2012,KHDR},
including dissipation and thermal noise for the cavity and mechanical
modes. A time dependent optomechanical interaction $g\left(t\right)$
\cite{quantum_memory_He2009} allows the entangled modes to transfer
to and from the mechanical modes $\hat{b}_{1}$ and $\hat{b}_{2}$.
The output fields are then shaped optimally to match the $\sech$-shaped
inputs, for maximum retrieval efficiency \cite{quantum_memory_He2009,sech_pulse_He2009},
with subsequent measurement of the stored entanglement. 

\paragraph{Time-dependent coupling and decay}

The optomechanical systems are modeled using the standard single mode
theory \cite{Hamil_Brag,Hamil_pierre,Pace-Collett-Walls}, following
techniques explained in previous papers \cite{KHDR}. It is convenient
to introduce a dimensionless time variable, $\tau=\Gamma_{c}t$, relative
to the optomechanical cavity decay rate $\Gamma_{c}$. To obtain universally
valid results covering a range of different cases, all other times,
frequencies and couplings are given in dimensionless units with derivatives
$\dot{f}\equiv\partial f/\partial\tau$. 

The equations of motion for the source cavity modes $a_{1},a_{2}$,
are obtained in the absence of thermal noise, assuming the only losses
are due to input/output coupling with a transmissivity $\kappa\left(\tau\right)$.
Using input-output theory, with inputs $a_{k,\text{in}}$ and outputs
$a_{k,\text{out}}$ one obtains \cite{inout-Gardiner-Collett,inout_Yurke}:
\begin{eqnarray}
\dot{a}_{k} & = & -\kappa\left(\tau\right)a_{k}+\sqrt{2\kappa\left(\tau\right)}a_{k,\text{in}}\nonumber \\
a_{k,\text{out}} & \equiv & \sqrt{2\kappa\left(\tau\right)}a_{k}-a_{k,\text{in}}.\label{eq:basic input-output}
\end{eqnarray}
We wish to generate a sech-shaped output pulse, $a_{out}\propto\sech\left(\tau\right)$.
This is achieved using a dimensionless mirror transmissivity defined
according to $\kappa\left(\tau\right)=\left[1+\tanh\left(\tau\right)\right]/2.$
 We solve for Eq. (\ref{eq:basic input-output}) , giving
\begin{eqnarray}
a_{k} & = & a_{k}\left(-\infty\right)\sqrt{\left[\frac{1-\tanh\left(\tau\right)}{2}\right]}+a_{vac}\nonumber \\
a_{k,out} & = & \frac{a\left(-\infty\right)}{\sqrt{2}}\sech\left(\tau\right)+a'_{vac}\,.\label{eq:source_solution}
\end{eqnarray}

The operators $a_{\text{vac}}$, $a'_{\text{vac}}$ are the source
input and output vacuum noises respectively. The cavities are cascaded,
so $d_{k,\text{in}}=a_{k,\text{out}}$, and from the input-output
relations, $d_{k,\text{out}}\equiv\sqrt{2}d_{k}-d_{k,\text{in}}.$
The optomechanical systems satisfy the standard quantum Langevin equations
\cite{Pace-Collett-Walls,KHDR} with cavity detuning $\delta\omega$,
mechanical loss $\gamma_{m}$, and optomechanical coupling $\chi$,
in dimensionless units. 

Assuming an intense red-detuned pump with $\delta\omega=\omega_{m}$,
and a resulting adiabatic coupling of $g=i\chi E/\left(\Gamma_{c}+i\delta\omega\right)$,
the linearized Hamiltonian for cavities $1$ and $2$ is $\hat{H}_{a,k}/\hbar\approx i\left(g^{*}d_{k}b_{k}^{\dagger}-gd_{k}^{\dagger}b_{k}\right)$.
Here, $d_{k}$ is a small fluctuation around the steady state in a
frame rotating with detuning $\delta\omega$. We determine the time
dependence of the optomechanical interaction strengths $g\left(\tau\right)$
of cavities $1$ and $2$, using previous work on quantum memories
\cite{sech_pulse_He2009}. 

To understand the mode-matching method, we start by analysing the
linearized equations without losses in the mechanical oscillator,
and without vacuum noise terms. These will be included in the full
numerical analysis, given next. At this stage, we have that:
\begin{eqnarray}
\dot{d}_{k} & = & -d_{k}-ig\left(\tau\right)b_{k}+\sqrt{2}d_{k,\text{in}}\nonumber \\
\dot{b}_{k} & = & -ig\left(\tau\right)d_{k}\,.
\end{eqnarray}
To find conditions for perfect input coupling, we require that $d_{\text{out}}=0$
in the absence of vacuum noise. Hence $d_{k,\text{in}}=\sqrt{2}d_{k}$,
leading to $\dot{d}_{k}=d_{k}-ig\left(\tau\right)b_{k}$. If we further
assume $b_{-\infty}=0$, again neglecting vacuum noise, then it follows
that $-ig\left(\tau\right)=\dot{b_{k}}/d_{k}$, giving
\begin{eqnarray}
\left(\dot{d_{k}}+igb_{k}\right)/d_{k} & = & \dot{d_{k}}/d_{k}-\left(\dot{b}^{2}\right)/\left(2d_{k}^{2}\right)=1\,.\label{eq:eqn_motion1}
\end{eqnarray}
Now we note that $d_{k}=a\left(-\infty\right)\sech\left(\tau\right)/2$,
and solving Eq. (\ref{eq:eqn_motion1}) gives us $b_{k}=ia\left(-\infty\right)\left[1+\tanh\left(\tau\right)\right]/2$.
From $-ig\left(\tau\right)=\dot{b_{k}}/d_{k}$, we obtain the input
modulation requirement of $g\left(\tau\right)=-\sech\left(\tau-\tau_{1}\right)$,
where $\tau_{1}$ is the peak transmission of the input. The output
modulation is identical apart from a shifted time-origin, from the
symmetry of the input/output relations under interchange of the input
and output terms.  \textcolor{green}{}

\paragraph{Output modes $\hat{A}_{1},\hat{A}_{2}$}

Detecting the stored entanglement requires an output measurement on
temporal modes ${\displaystyle \hat{A}_{k}=\int_{-\infty}^{\infty}u_{k}\left(\tau'\right)\hat{d}_{\text{out}}\left(\tau'\right)d\tau'}$
such that $\left[\hat{A}_{k},\hat{A}_{k}^{\dagger}\right]=1$ \cite{ent-opto-Hofer}.
We can then observe entanglement between $\hat{A}_{1}$ and $\hat{A}_{2}$
on a scale comparable with the initial entanglement between $\hat{a}_{1}$
and $\hat{a}_{2}$. Choosing the output pulse to be an identical shape
to the input, so that $\hat{a}_{k,\text{out}}\propto\sech\left(\tau\right),$
we have $u_{k}\left(\tau\right)=u\left(\tau\right)=N\cdot\sech\left(\tau\right)$.
This leads to a normalization of
\begin{equation}
{\displaystyle N=1/\sqrt{\int_{-\infty}^{\infty}\sech\left(\tau\right)^{2}d\tau}=\sqrt{\frac{1}{2}}}\,,
\end{equation}
The normalization constant for a restricted time-domain can also be
found, which leads to minor corrections.

\begin{figure}
\centering{}\includegraphics[width=0.8\columnwidth]{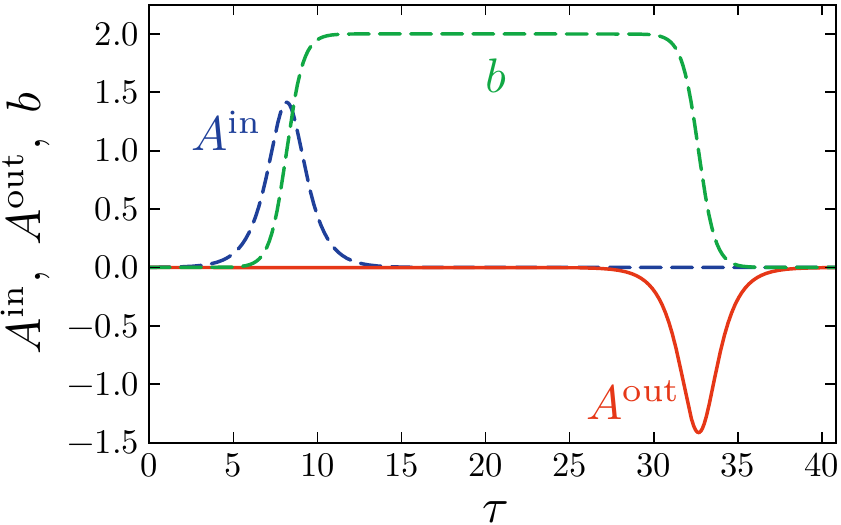}\caption{Temporal behaviour of the input field $a_{in}$, the output field
$A_{out}$ and the mechanical state $b$.\label{fig:Temporal-behaviour-of}}
\end{figure}

\begin{figure}
\centering{}\includegraphics[width=0.8\columnwidth]{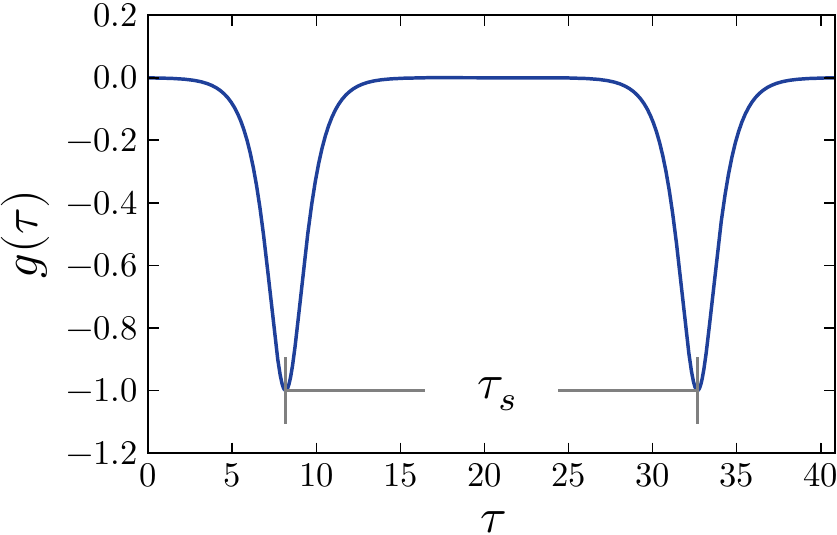}\caption{Temporal behaviour of the coupling strength.\label{fig:Temporal-behaviour-of-coupling} }
\end{figure}

\paragraph{Wigner representation and stochastic equations}

There is thermal noise in the mechanical mode due to the interaction
with its reservoir. In order to calculate these effects, and to include
vacuum noise terms rigorously, it is useful to introduce a Wigner
representation of the system density matrix \cite{Wigner,Wigner_Castin,Wigner_Steel}.
The Wigner function for the initial entangled state is given by \cite{Walls_book}
\begin{equation}
{\displaystyle W\left(\alpha_{+},\alpha_{-},\tau_{0}\right)=\frac{4}{\pi^{2}}\exp\left[-2\left(\frac{\left|\alpha_{+}\right|^{2}}{e^{2r}}+\frac{\left|\alpha_{-}\right|^{2}}{e^{-2r}}\right)\right]\,,}
\end{equation}
where $\alpha_{\pm}{\displaystyle =\left(\alpha_{1}\pm\alpha_{2}^{*}\right)/\sqrt{2}}$
and $r$ is the squeezing parameter that characterizes the degree
of entanglement. One can sample $\alpha_{1},\alpha_{2}$ by generating
Gaussian noise vectors $\xi_{x}^{\pm},\xi_{y}^{\pm}$ with unit variance,
defining $\alpha_{\pm}{\displaystyle =\left[\xi_{x}^{\pm}+i\xi_{y}^{\pm}\right]e^{\pm r}/2}$
and then obtaining mode amplitudes ${\displaystyle \alpha_{1}}=\left(\alpha_{+}+\alpha_{-}\right)/\sqrt{2}$
and ${\displaystyle \alpha_{2}}=\left(\alpha_{+}^{*}-\alpha_{-}^{*}\right)/\sqrt{2}$. 

Quantum dynamical time evolution now follows a stochastic equation,
after truncating third order derivative terms which are negligible
for large occupation numbers. It is also possible to use a positive-P
representation, which requires no truncation \cite{KHDR}. Taking
account of the cascaded input-output relations, the coupled equations
describing time evolution of the Wigner amplitudes for the entangled
source cavities $\alpha_{k}$, optical cavities $\delta_{k}$ and
mechanical modes $\beta_{k}$ are given, for $k=1,2$, by
\begin{eqnarray}
\dot{\alpha}_{k} & = & -\kappa\left(\tau\right)\alpha_{k}+\sqrt{2\kappa\left(\tau\right)}\xi_{k}\nonumber \\
\dot{\delta}_{k} & = & -\delta_{k}-ig\left(\tau\right)\beta_{k}+2\sqrt{\kappa\left(\tau\right)}\alpha_{k}-\sqrt{2}\xi_{k}\nonumber \\
\dot{\beta}_{k} & = & -\gamma_{m}\beta_{k}-ig\left(\tau\right)\delta_{k}+\sqrt{2\gamma_{m}\left(2\bar{n}_{\text{th},m}+1\right)}\xi_{2+k}\,.\label{eq:eq_system}
\end{eqnarray}
Here $\bar{n}_{\text{th},m}=1/\left[\exp\left(\hbar\Gamma_{c}\omega_{m}/k_{B}T\right)-1\right]$
is the average phonon number in the mechanical bath, and $\xi_{k}$
are complex Gaussian noises with variances that correspond to the
'half-quanta' occupations of symmetric Wigner vacuum correlations,
$\langle\xi_{k}\left(\tau\right)\xi_{l}^{*}\left(\tau'\right)\rangle=\frac{1}{2}\delta_{kl}\delta\left(\tau-\tau'\right)$.
Using the input-output relations again, we obtain the expression $\delta_{k,\text{out}}=\sqrt{2}\delta_{k}-\sqrt{2\kappa\left(\tau\right)}\alpha_{k}+\xi_{k}$.
The output modes  used for detecting entanglement are then:
\begin{align}
{\displaystyle A_{k,\text{out}}} & =\int_{\tau_{1}+\tau_{s}/2}^{\tau_{\text{max}}}u\left(\tau-\tau_{2}\right)\\
 & \,\,\,\,\times\left(\left[\sqrt{2}\delta_{k}-\sqrt{2\kappa\left(\tau\right)}\alpha_{k}\right]+\xi_{k}\right)d\tau\,.\nonumber 
\end{align}
Note that the time integration for the output modes only starts after
the first transfer pulse has been completed. 

\paragraph{Experimental parameters}

We assume that the optical modes of cavities $1$ and $2$ are initially
in a vacuum state. The source cavity and cavities $1$ and $2$ are
connected by a perfect, lossless waveguide. There is only one source
of decoherence affecting the optical cavity modes, which is thermal
noise in the mechanical modes. 

Our simulations used experimental parameter values very similar to
the optomechanical experiment values reported by Chan et. al. \cite{cooling_Chan_2011_Nature}.
The mechanical modes have an initial occupation of $n_{th,b}\left(0\right)=0.7$,
corresponding to a reservoir temperature of $200\,\text{mK}$. The
cavity decay rate is $\Gamma_{c}/2\pi=0.26\ \text{GHz}$. Relative
to this time-scale, the mechanical oscillator has dimensionless resonance
frequency $\omega_{m}/2\pi=14.23$, with a mechanical dissipation
rate of $\gamma_{m}/2\pi=1.59\cdot10{}^{-5}$ and an optomechanical
coupling strength of $\chi_{0}/2\pi=3.5\times10^{-3}$, which justifies
the linearization \cite{ent-opto-Giovannetti,ent-opto-Mancini} and
adiabatic approximations \cite{ent-opto-Hofer}. 

The time dependent source cavity decay rate that shapes the entangled
modes is given by ${\displaystyle \kappa\left(\tau\right)=\frac{1}{2}\left[1+\tanh\left(\tau-\tau_{1}\right)\right]}$,
while the effective coupling strength is
\begin{eqnarray}
g\left(\tau\right) & =\begin{cases}
 & -\sqrt{2}u\left(\tau-\tau_{1}\right)\,,\forall\,0\leq\tau\leq\tau_{1}+\frac{\tau_{s}}{2}\\
 & -\sqrt{2}u\left(\tau-\tau_{2}\right)\,,\forall\,\tau_{1}+\frac{\tau_{s}}{2}\leq\tau\leq\tau_{\text{max}}\,,
\end{cases}\label{eq:g(t)}
\end{eqnarray}
where $\tau_{1}=8.17$ and $\tau_{2}=\tau_{1}+\tau_{s}$ are the dimensionless
times when the storing and reading pulses peak, and $\tau_{\text{max}}=2\tau_{1}+\tau_{s}$,
while $\tau_{s}$ is the dimensionless time between the peaks of the
storage and readout pulses. It is also the storage time of the entangled
state in the mechanical mode, as illustrated in Fig. \ref{fig:Temporal-behaviour-of-coupling}.

\begin{figure}
\begin{centering}
\includegraphics[width=0.8\columnwidth]{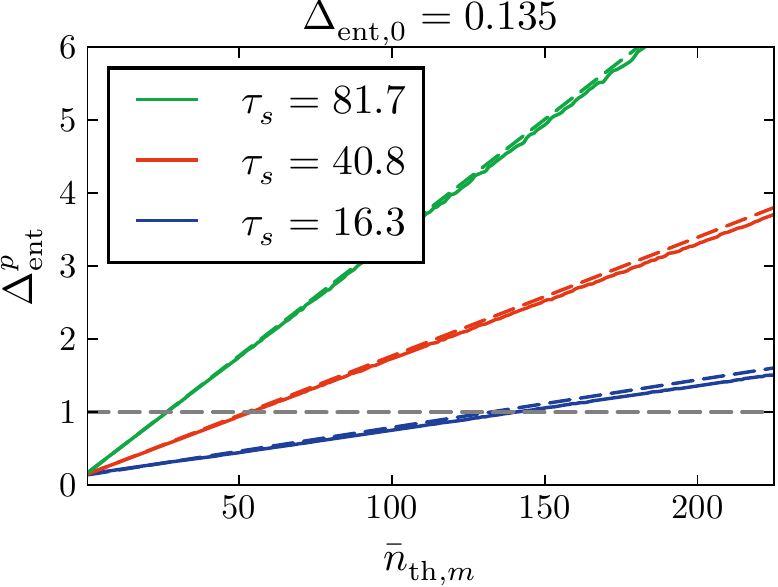}
\par\end{centering}
\medskip{}

\begin{centering}
\includegraphics[width=0.8\columnwidth]{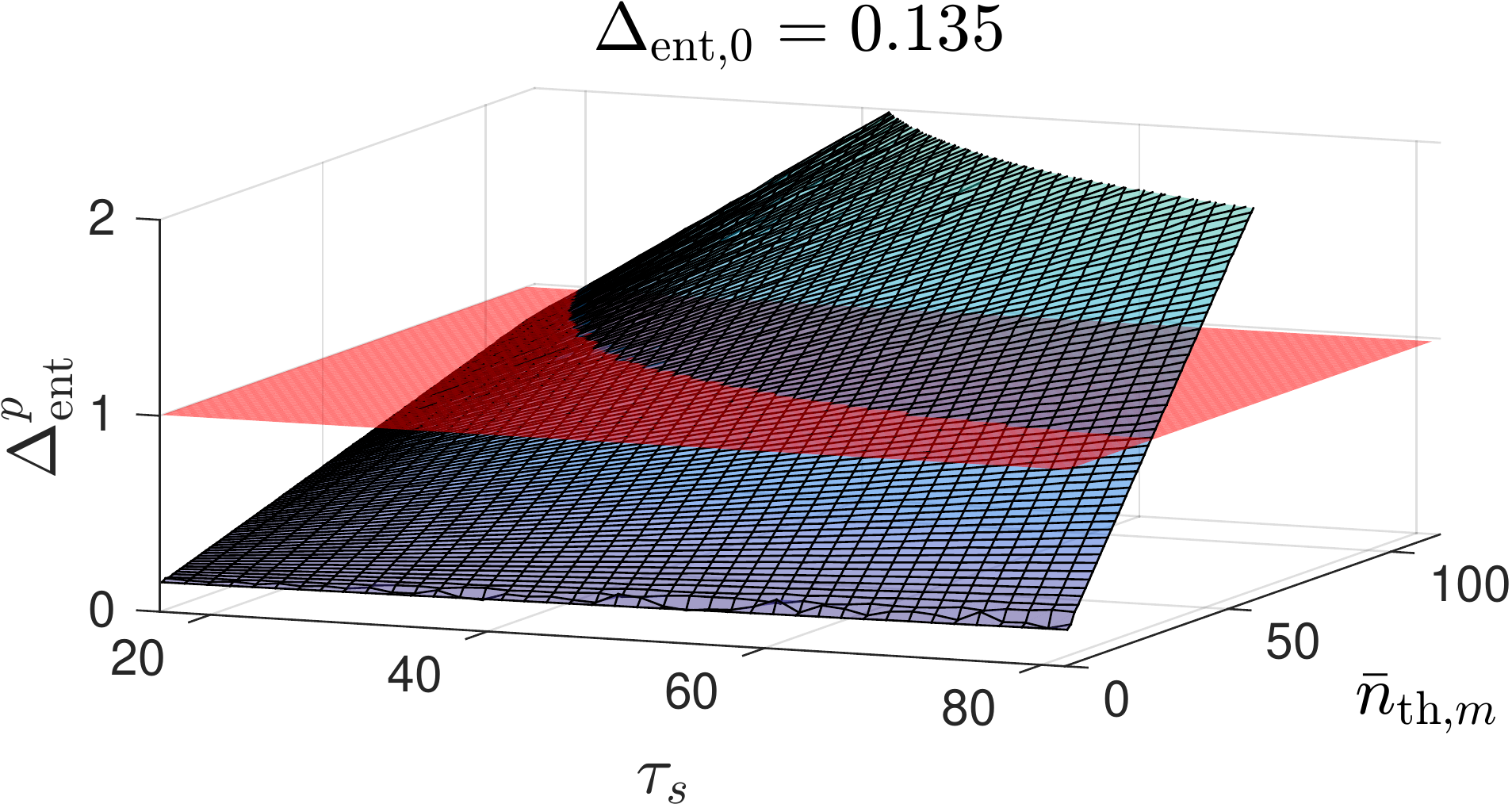}
\par\end{centering}
\caption{Entanglement as a function of temperature and storage time. The squeezing
parameter is $r=1$, characterizing the degree of entanglement in
the source cavity, while $\Delta_{\text{ent},0}=0.135$ is the initial
value of entanglement. \textcolor{red}{\label{fig:Entanglement-vs-time}}}
\end{figure}

\paragraph{Entanglement criterion}

We use the phase- and gain-optimized product signature as an entanglement
criterion \cite{entanglement-criteria-Giovanetti2003}, defined as:
\begin{equation}
{\displaystyle \Delta_{ent}^{p}=\frac{4\Delta\left(X_{1}-GX_{2}^{\theta}\right)\Delta\left(P_{1}+GP_{2}^{\theta}\right)}{\left(1+G^{2}\right)}<1}\,,\label{eq:ent_criterion}
\end{equation}
where ${\displaystyle X_{k}^{\theta}}=\frac{1}{2}\left[e^{-i\theta}A_{k,\text{out}}+e^{i\theta}A_{k,\text{out}}^{\dagger}\right]$,
$P_{k}^{\theta}=X_{k}^{\theta+\pi/2}$ and $G$ is an adjustable real
constant. In particular, ${\displaystyle X_{k}=X_{k}^{0}}$, ${\displaystyle P_{k}=P_{k}^{0}}$
are the usual phase and amplitude quadratures. We minimize $\Delta_{\text{ent}}^{p}$
with respect to the gain $G$ and phase $\theta$ simultaneously.
When inequality (\ref{eq:ent_criterion}) holds, the optimised value
of $\Delta_{\text{ent}}^{p}$ characterizes the degree of quantum
entanglement between the modes \cite{entmeasPPT}. 

We compute $\Delta_{\text{ent}}^{p}$ in Eq. (\ref{eq:ent_criterion})
as a function of thermal reservoir occupation number for a set of
different storage times and a fixed squeezing parameter. To give an
approximate analytic prediction, we consider only the degradation
of the entanglement during its storage period in the mechanical oscillators.
Using results described in \cite{epr_decoherence}, we predict an
entanglement value of
\begin{align}
\Delta_{\text{ent}}^{p} & =e^{-2\gamma_{m}\tau_{s}}e^{-2r}+\left(1-e^{-2\gamma_{m}\tau_{s}}\right)\left(1+2\bar{n}_{\text{th},m}\right)\,.
\end{align}
Fig. (\ref{fig:Entanglement-vs-time}) shows the predicted entanglement
results for squeezing parameter $r=1$ and three different storage
times $\tau_{s}=16.3,\ 40.8,\ 81.7$, corresponding to $10\textrm{\,ns}$,
$25\,\textrm{ns}$ and $50\,\textrm{ns}$, respectively. The solid
lines indicate simulation results and dashed lines theoretical predictions. 

The simulation results were obtained by solving Eqs. (\ref{eq:eq_system})
with a stochastic $4$th order Runge-Kutta algorithm, $3000$ time-steps
and $\approx2\cdot10^{8}$ samples, using open-source software \cite{Kiesewetter}.
They are in good agreement with our analytic predictions, in fact
exhibiting slightly more favorable entanglement. A larger initial
entanglement in the source cavity and a shorter storage time gives
even better output temporal mode entanglement. 

\paragraph{Quantum fidelity}

We consider the quantum fidelity measure $\mathcal{F}=\langle\psi|\rho|\psi\rangle\,$,
where $|\psi\rangle$ is the two mode squeezed state and $\rho$ is
the density operator describing the temporal output modes. The fidelity
quantifies the efficiency of our entanglement protocol as the entanglement
in output temporal modes rely on successful entangled state transfer
from the source cavity. In the Wigner representation \cite{fidelity_cahill_1969,fidelity_schumacher1996},
\begin{equation}
\mathcal{F}=\pi^{2}\int\ W_{\psi}\left(\alpha_{1},\alpha_{2}\right)W_{\rho}\left(\alpha_{1},\alpha_{2}\right)\ d^{2}\alpha_{1}d^{2}\alpha_{2}\,.\label{eq:quantum_fidelity}
\end{equation}
From the quantum simulations, we obtain sampled temporal output modes
from the Wigner function $W_{\rho}$. The quantum fidelity $\mathcal{F}$
is then computed using
\begin{eqnarray}
\mathcal{F} & = & \frac{\pi^{2}}{N_{\text{sample}}}\sum_{i}W_{\psi}\left(A_{1,\text{out}}^{i},A_{2,\text{out}}^{i}\right)\,,\label{eq:sampled_fidelity}
\end{eqnarray}
where $A_{k,\text{out}}^{i}$ is the $i$th sample of temporal output
mode $A_{k,\text{out}}$ and $N_{\text{sample}}$ is the total number
of samples taken.

\begin{figure}
\begin{centering}
\includegraphics[width=0.8\columnwidth]{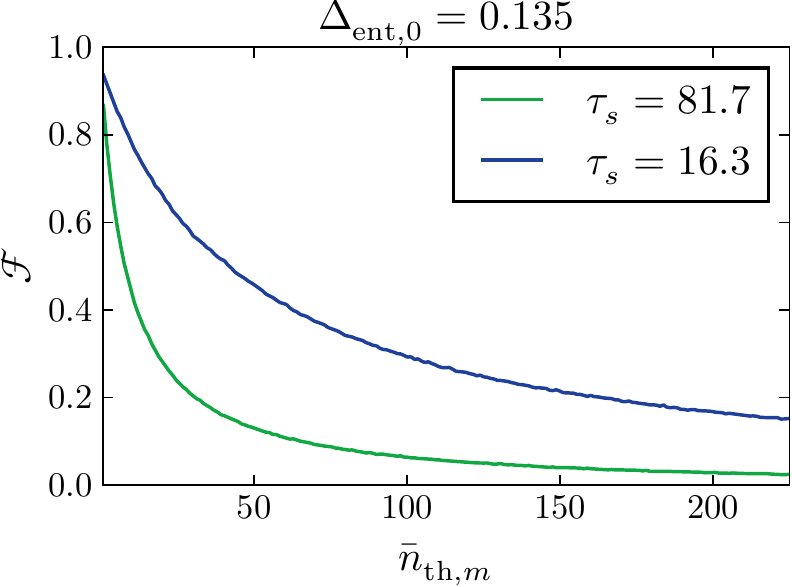}
\par\end{centering}
\medskip{}

\begin{centering}
\includegraphics[width=0.8\columnwidth]{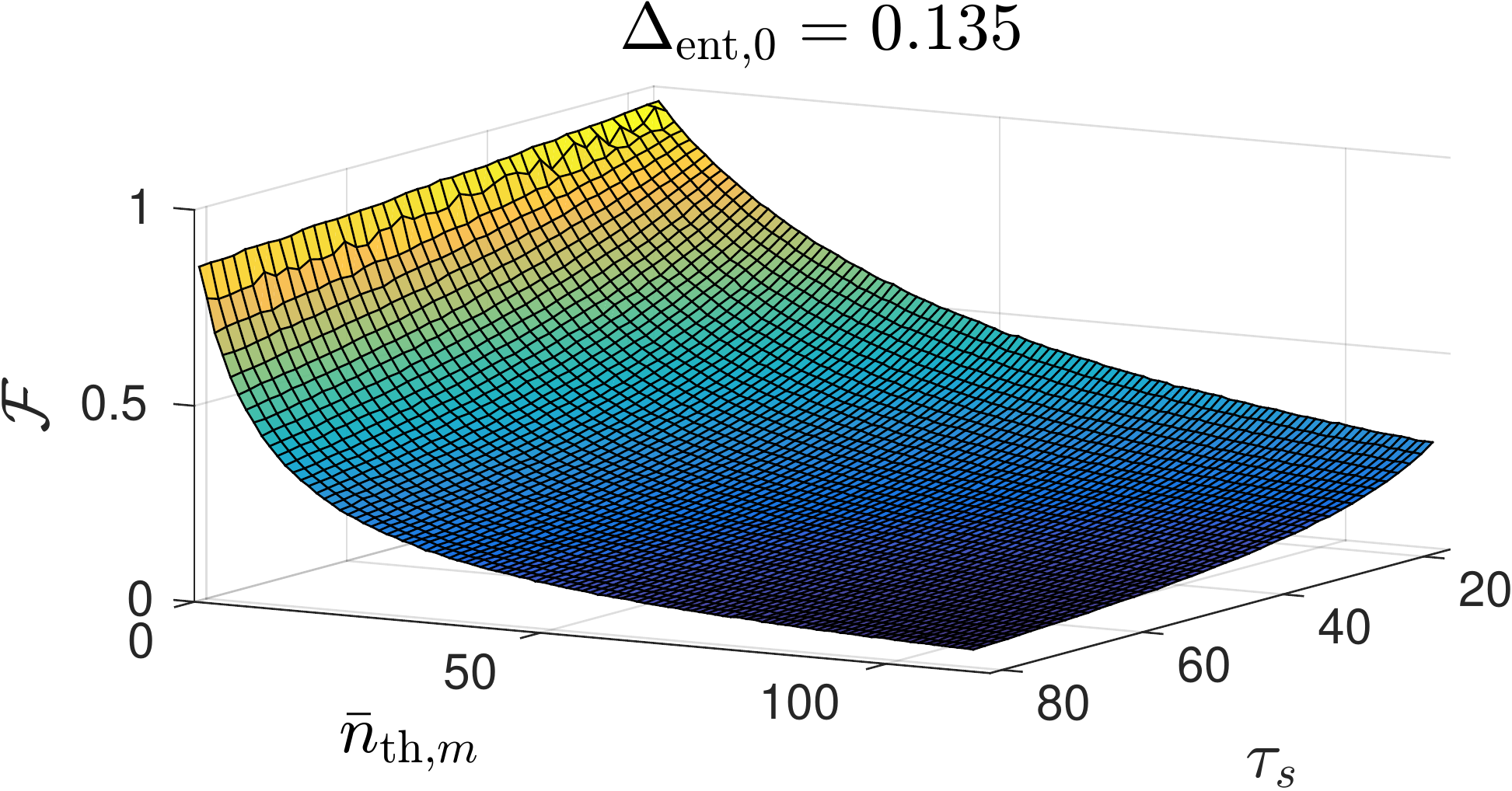}
\par\end{centering}
\caption{Fidelity $\mathcal{F}$ as a function of the thermal bath occupation
number and storage time. Other parameters as in Fig \ref{fig:Entanglement-vs-time}.
\label{fig:fidelity}}
\end{figure}

The quantum fidelity in Eq. (\ref{eq:sampled_fidelity}) is also computed
as a function of reservoir temperature and storage time. The top plot
in Fig. (\ref{fig:fidelity}) shows the steep drop in fidelity as
storage time is increased. Comparing plots in Fig. (\ref{fig:Entanglement-vs-time})
and Fig. (\ref{fig:fidelity}) shows that a fidelity $\mathcal{F}$
of at least about $0.3$ is needed for entangled output modes.

\paragraph{EPR-steering}

\begin{figure}
\begin{centering}
\includegraphics[width=0.8\columnwidth]{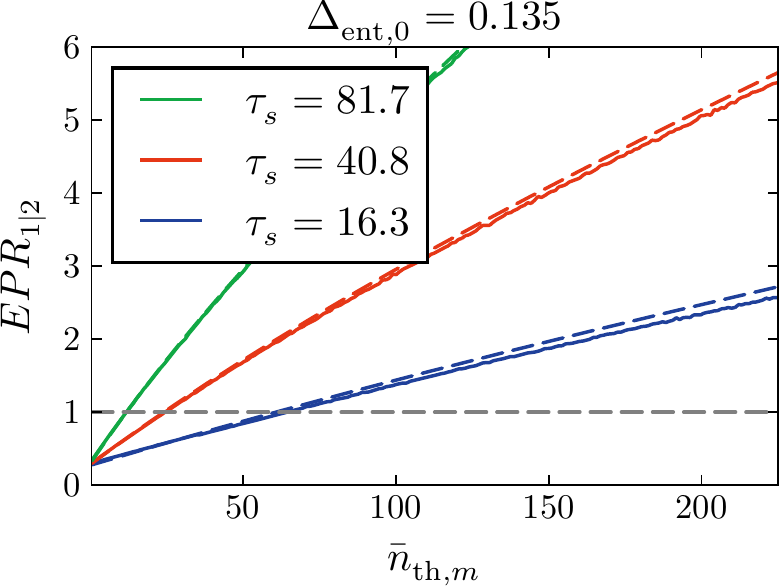}
\par\end{centering}
\caption{\label{fig:EPR-steering}EPR-steering as a function of the thermal
bath occupation number for three different storage times. Other parameters
as in Fig \ref{fig:Entanglement-vs-time}.}
\end{figure}

In addition to entanglement, we also analyze the stronger, asymmetric
nonlocality signature known as the EPR-steering that links directly
to the EPR paradox \cite{Reid1989,EPR1935,Wisemansteering}. We use
the CV signature for steering of system $1$ by system $2$ \cite{Reid1989}
\begin{eqnarray}
EPR_{1|2} & = & 4\Delta\left(X_{1}-GX_{2}^{\theta}\right)\Delta\left(P_{1}+GP_{2}^{\theta}\right)<1\,,
\end{eqnarray}
with $X,\:P$ as previously and an optimized gain $G$. Fig. \ref{fig:EPR-steering}
shows the predicted results for EPR-steering. The solid lines indicate
simulation results and the dashed lines give analytic predictions.
The analytic predictions were obtained analogously to the entanglement
predictions. Using the results described in \cite{epr_decoherence},
we obtain
\begin{align}
EPR_{1|2} & =\frac{2ab\left(1-b\right)c+b^{2}+c^{2}\left(1-b\right)^{2}}{ab+\left(1-b\right)c}\,,\label{eq:steering}
\end{align}
where $a\equiv\cosh\left(2r\right),\ b\equiv e^{-2\gamma_{m}\tau_{s}},\ c\equiv\left(1+2\bar{n}_{\text{th},m}\right)\,$.
Because of the symmetric setup, $EPR_{1|2}$ and $EPR_{2|1}$ are
equal in magnitude.

\paragraph{Conclusions}

In summary, our results show that a synchronous pulsed experiment
can, in principle, transfer, store and read out macroscopic entanglement
of two mechanical oscillators with nearly $100\%$ efficiency under
ideal conditions. Due to finite temperature effects and damping, this
effect is degraded in a predictable way. We calculate the quantitative
effects of known decoherence on this proposed experiment. The experimental
objectives would be to place a bound on additional decoherence that
may occur due to the oscillator separation, to test the validity of
Furry's hypothesis.
\begin{acknowledgments}
This research was supported in part by the National Science Foundation
under Grant NSF PHY-1125915, and by the Australian Research Council
under Grant DP140104584.
\end{acknowledgments}

\end{document}